\documentclass[aip,reprint]{revtex4-1}
\usepackage{graphics,graphicx,dcolumn,bm,fleqn,epic,eepic,float}
\usepackage{amssymb,amsmath,multirow,rotate,color,float}
\usepackage[latin1]{inputenc}

\usepackage[dvips]{epsfig}

\begin{document}

\title[Regimes of Wetting Transitions on Superhydrophobic Textures]{Regimes of Wetting Transitions on Superhydrophobic Textures Conditioned by Energy of Receding Contact Lines}

\author{Alexander L. Dubov}
\affiliation{A.N.~Frumkin Institute of Physical Chemistry and Electrochemistry, Russian Academy of Sciences, 31 Leninsky
Prospect, 119071 Moscow, Russia}
\affiliation{DWI - Leibniz Institute for Interactive Materials, RWTH Aachen, Forckenbeckstrasse 50, 52056 Aachen, Germany}

\author{Ahmed Mourran}
\affiliation{DWI - Leibniz Institute for Interactive Materials, RWTH Aachen, Forckenbeckstrasse 50, 52056 Aachen, Germany}

\author{Martin M\"oller}
\affiliation{DWI - Leibniz Institute for Interactive Materials, RWTH Aachen, Forckenbeckstrasse 50, 52056 Aachen, Germany}

\author{Olga I. Vinogradova}
\email{oivinograd@yahoo.com}

\affiliation{A.N.~Frumkin Institute of Physical Chemistry and Electrochemistry, Russian Academy of Sciences, 31 Leninsky
Prospect, 119071 Moscow, Russia}
\affiliation{DWI - Leibniz Institute for Interactive Materials, RWTH Aachen, Forckenbeckstrasse 50, 52056 Aachen, Germany}
\affiliation{Department of Physics, M.V.~Lomonosov Moscow State University, 119991 Moscow, Russia}

%\pacs{68.08.-p}{Liquid-solid interfaces}
%\pacs{68.08.Bc}{Wetting}

\begin{abstract}
We discuss an evaporation-induced wetting transition  on superhydrophobic stripes, and show that depending on the elastic energy of the deformed
contact line, which determines the value of an instantaneous apparent contact angle, two different scenarios occur. For relatively dilute stripes the receding angle is above 90$^\circ$, and the sudden impalement transition  happens  due to an increase of a curvature of an evaporating drop. For dense stripes the slow impregnation transition  commences when the apparent angle reaches 90$^\circ$ and represents the impregnation of the grooves from the triple contact line towards the drop center.
\end{abstract}

\maketitle

%\section{Introduction}

\textbf{Introduction.~--}
On a superhydrophobic (SH) surface, which is both rough and hydrophobic,  the contact angle of water is very large,~\cite{quere.d:2005} and two states are possible for a drop. In the Wenzel state water penetrates the texture completely,~\cite{wenzel.rn:1936} but in the Cassie state  air is trapped by texture, so that liquid sits on the top of the asperities.~\cite{Cassie}
In most applications (microfluidics, ``self-cleaning'', water-repellent materials,   bio-medical devices) the Cassie state is desired, which, however, could be unstable or metastable, so that irreversible transitions
towards the more stable Wenzel state can be provoked.~\cite{sbragaglia.m:2007,bormashenko:2007_2,pirat.c:2008,reyssat.m:2008}

During the last decade, much insight has been gained into the relative stability of
the Cassie state, the Cassie-to-Wenzel wetting transition.
Several criteria to determine a preferable state have been suggested.~\cite{quere.d:2005,cottin-bizonne2004}
If the liquid/air interfaces are flat (very large drops), then for a texture with a given roughness $r$ (ratio between the actual surface area of a rough solid and its projected surface area) and solid-liquid fraction, $\phi_{S}$, Cassie and Wenzel angles are equal for a material with Young's angle  $\theta_c$ defined as~\cite{bico.j:2002}

\begin{equation}
\cos\theta_c=\frac{\phi_{S}-1}{\emph{r}-\phi_{S}}.
\label{thitac}
\end{equation}

The Cassie state should be stable if $\theta>\theta_{c}$, while the Wenzel state is preferred at $\theta<\theta_{c}$. Hence, no stable Cassie state is possible on hydrophilic textures. These relation, however, becomes invalid for small drops due to a Laplace pressure, $\Delta\emph{p} = 2\gamma/\emph{R}$, where $\gamma$ is the surface tension  and $\emph{R}$ is the drop radius. Accounting for this pressure difference gives a more general thermodynamic criterion for the stability of the Cassie state:~\cite{cottin-bizonne2004} $(\gamma+\emph{h}\Delta\emph{p})\cos\theta_{c}-\gamma \cos\theta>0$, where $h$ is the depth of the texture.
 For low $\phi_S$ these criteria deny the observed experimentally Cassie state suggesting it is metastable. To interprete these several models have been proposed,~\cite{liu.b:2006,reyssat.m:2008,Giacomello:2012} such as a touching the bottom side of the texture~\cite{reyssat.m:2008} or reaching the local advancing contact angle on a vertical side of the texture.~\cite{dubov:2013,Tuteja2008}

 % \textbf{Compare and contrast the paper to other literature. Demonstrate why the results and are new, interesting, and/or surprising?}

 Most investigations of a wetting transition have been conducted with a dilute array of SH pillars due to their remarkable  wetting  properties.  In this letter we study the wetting transition on textures decorated by highly anisotropic SH grooves and vary $\phi_S$ in a very large range.
 %\textbf{Why these textures were rarely studied before and what was already done? Are they interesting for wetting purposes?}
 Such textures, which do not necessarily provide large apparent contact angles, now become extremely important in the context of microfluidic applications.~\cite{vinogradova2012} For instance, dilute SH stripes show the remarkable drag-reducing ability.~\cite{vinogradova.oi:2011,tsai2009} Dense SH grooves  generate transverse hydrodynamic phenomena~\cite{feuillebois.f:2010b} and can be successfully used to separate tiny particles~\cite{pimponi.d:2014,asmolov.es:2015} or enhance mixing rate~\cite{ou.j:2007,nizkaya.tv:2015} in microfluidic devices. These striped textures amplify electrokinetic pumping~\cite{bahga:2009,belyaev.av:2011a} and are employed for sorting droplets.~\cite{nilsson.ma:2011,nilsson.ma:2012,sbragaglia.m:2014}

We study a wetting transition by monitoring the evaporation of a drop. An evaporating
drop is known to produce various intriguing phenomena, such as coffee stains~\cite{deegan.rd:2000} or wine tears,~\cite{hosoi.a:2001}
and has been already studied with isotropic SH surfaces,~\cite{reyssat.m:2008,mchale.g:2005,tsai.p:2010b} although the quantitative understanding is still at its infancy.
During the evaporation the
drop becomes smaller and the contact angle retracts. At some moment the transition to the Wenzel state occurs.

Some recent results may add a new dimension to the problem of evaporation-induced wetting transition. It has been shown~\cite{dubov:2014_2,Reyssat2009CAH} that for relatively low $\phi_S$, the receding (i.e. the smallest possible) angle, $\cos\theta_r^{\ast}$,  on the SH surface, which is
above $90^\circ$,
 %and always isotropic~\cite{dorrer.c:2006,Choi2009CAH,Gauthier2013},
 depends significantly on the elastic energy of the deformed contact line. During receding the contact line is bent at the gas areas being pinned at solid areas. The value of a receding angle on a striped SH surface can then be evaluated as:~\cite{dubov:2014_2}
\begin{equation}
\cos\theta_r^{\ast}= - 1 + \phi_{S} (1 + \cos\theta_{r})-\frac{2a+1}{2}\phi_S^2\ln\phi_S,
\label{eq_rec}
\end{equation}
where $\theta_{r}$ is the receding angle of the material. The nonlinear (logarithmic)
term here includes the parameter $a$, which reflects the relative length of a side part of the pinned contour. This term represents an elastic contribution to $\cos\theta_r^{\ast}$, which has been found to be large even at moderate $\phi_S$. It is therefore natural to assume that an elastic distortion of the contact line could be important in determining the onset of the evaporation-induced wetting transition. All existing approaches neglect the contribution of the triple contact line to the energy of the system (except some qualitative free energy arguments~\cite{bormashenko:2012,Chen2012}) and we are unaware of any prior work that has addressed
this issue.

%\textbf{State the main results of the paper, and outline how these results can be used in various applications. }

The aim of our letter is to explore experimentally and to discuss quantitatively the possible effect of
the contact line elasticity on the wetting transitions.  We show here that it is generally neither small nor negligible, and  that depending on the degree of elastic deformation of the contact line two physically different scenarios, fast impalement and slow imbibition, could  occur.

%\section{Experimental}

\begin{figure}
  \begin{center}
  \includegraphics[width=6cm]{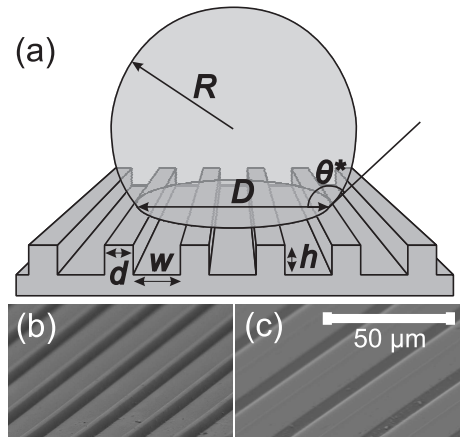}\\
  \caption{Sketch of droplet settled on the SH stripes (a) and microphotographs of superhydrophobic surfaces with $\phi_S = 0.49$ (b) and 0.33 (c) fabricated by a soft lithography}
  \label{fig_sketch}
  \end{center}
\end{figure}

\textbf{Experimental.~--}
SH rectangular grooves (see Fig.~\ref{fig_sketch}) of width $w$ separated by distance $d$ (both varied from 3 to 25~$\mu$m) and depth $h$ (5~$\mu$m) have been prepared to provide $\phi_S = d/(w+d)$ from 0.12 to 0.88 in the Cassie state. SH surfaces were made of polydimethylsiloxane by replication of silicon masters (Fig.~\ref{fig_sketch}(b,c)). The masters fabricated by projection lithography have been purchased from AMO Gmbh, Germany. They have been oxidized by plasma treatment in evacuated chamber with oxygen flow 30~mL/min and plasma power 200~W (PVA TePla 100 Plasma System), and coated with a trichloro(1H,1H,2H,2H-perfluorooctyl)silane (PFOTCS, 97\%, Sigma Aldrich) through the gas phase in evacuated dessicator at pressure 80~mbar for 3~hours. After removal of a bowl with PFOTCS from dessicator it was evacuated to full vacuum for 15~minutes in order to remove formed  clusters of PFOTCS from the silicon surface. Hydrophobized surfaces have been rinsed by isopropanol. Afterwards we fabricated hydrophobic PDMS replicas from silicon patterns by a common soft lithography~\cite{Xia1998} (Sylgard\circledR 184 Silicone elastomer KIT, Dow Corning). Flat surface of such PDMS showed a receding angle $\theta_{r}=84^\circ$ and advancing angle $\theta_{a}=115^\circ$.

SEM images were taken using a Hitachi S-4800 field emission scanning electron microscope. Texture geometry has been characterized by interference profilometry (WYKO NT2000, Veeco, USA) and optical microscopy (Axioplan 2 Imaging System, Zeiss, Germany).

To study CWT we monitored evaporation of milli-Q water droplet deposited onto a surface~\cite{mchale.g:2005,reyssat.m:2008} at room temperature and ambient atmospheric pressure. Measurement of evolution of droplet geometrical parameters, namely radius $R$, contact angle $\theta^\ast$ and base diameter $D$ (see Fig.~\ref{fig_sketch}), were performed with use of Drop Shape Analysis System DSA100 (Kr\"uss, Germany). Current SH wetting state of the drop was monitored by optical microscopy (Axioplan 2 Imaging System, Zeiss, Germany) equipped with a CCD camera recording up to 25 frames per second. The typical initial volume of the droplet for all types of measurements was 2~$\mu$L.

%\section{Results and discussion}

\begin{figure}[h]
  \centering
  \includegraphics[width=8cm]{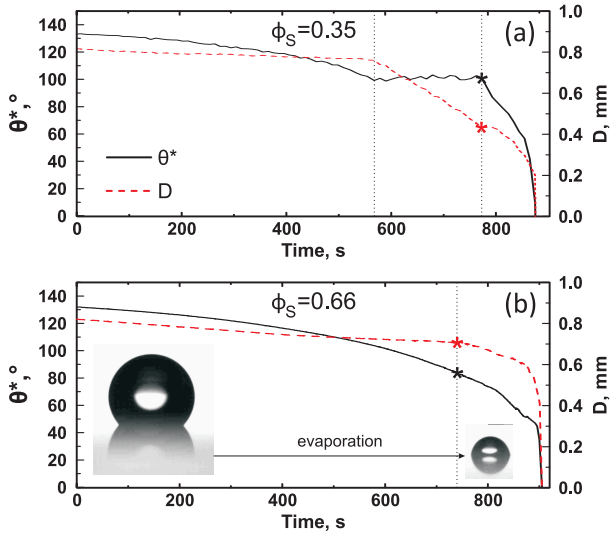}\\
  \caption{Typical dependence of an apparent contact angle (solid curve) and a drop base diameter $D$ (dashed curve) on time during evaporation for $\phi_S=0.35$~(a) and $\phi_S=0.66$ (b). The
symbols mark the transition points.}
  \label{fig_evap}
\end{figure}

\textbf{Results and discussion.~--}
During the evaporation the decrease of a droplet volume (and the increase of its curvature) is generally accompanied by a retraction of a triple contact line. Fig.~\ref{fig_evap} illustrates the typical changes of the instantaneous apparent contact angle and the base diameter during the evaporation. All measurements suggested that $\theta^\ast$ is practically isotropic despite the anisotropy of the texture.  We recall that observations of an isotropic receding angle for
a drop on striped surfaces have been reported before.~\cite{Choi2009CAH,dubov:2014_2}
%always practically isotropic (note that similar observations have been made before for different anisotropic textures~\cite{dorrer.c:2006,,Gauthier2013}) and
Note however that the transverse angle shows irregular oscillations of a period $w+d$, more pronounced for textures with larger $w$.

For relatively dilute patterns (see Fig.~\ref{fig_evap}(a)) an evaporation represents a three-stage process. On the first stage we observe a decrease in  $\theta^\ast$ at a nearly constant (due to a pinning of a contact line on defects) base diameter, $D$. During the second stage the contact angle reaches its minimum value, $\theta^\ast_r$ (above  $90^\circ$), and remains constant as the drop evaporates. The contact line becomes completely unpinned and slides fast. However, this state changes
when the drop becomes smaller than a critical radius, $R^\ast$. Then, the angle suddenly decreases, which implies a sharp increase in $D$. The contact line is very efficiently pinned,
and the angle gradually vanishes, from an initial value down to zero, despite the hydrophobicity of the material. This behavior results from an impalement of the drop in the texture.~\cite{reyssat.m:2008} Note that the impalement time is well below $40$~ms.
% \textbf{Compare with prior observations. What is the same? What is different?}
We remark that these observations are qualitatively similar to made before for dilute pillars.~\cite{bormashenko:2012,dubov:2013}

For dense patterns the evaporation occurs in two stages (see Fig.~\ref{fig_evap}(b)), and the first one is identical to observed for dilute patterns.  The evolution of a contact angle is smooth and does not indicate the difference between two steps of evaporation. However, the dependence of $D$ on time represents a piecewise function with a distinct fold point, which corresponds to $\theta^\ast \simeq 90^\circ$, where the second stage commences. Then, the base diameter starts to decrease much faster, and the contact angle vanishes.
This behavior can be interpreted as resulting from a wetting transition, which is much slower than for dilute patterns and obviously involves a completely different scenario we discuss below.

\begin{figure}[h]
  \centering
  \includegraphics[width=8cm]{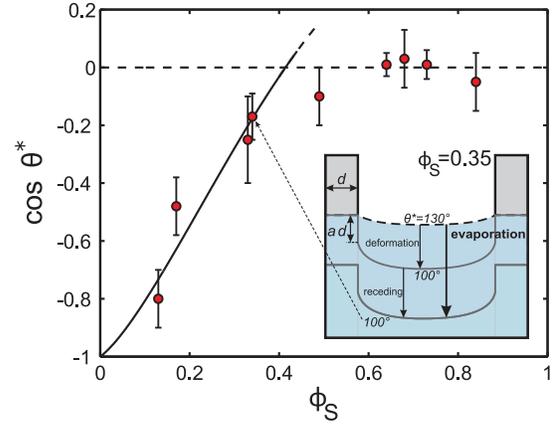}\\
  \caption{Effective contact angle of the wetting transition as a function of $\phi_S$. Red circles show experimental data, solid curve plots theoretical predictions (Eq.~\ref{eq_rec}), dashed line corresponds to 90$^\circ$. Inset shows (top view) sketch of a deformed contact line on the stripes with $\phi_S=0.35$}
  \label{fig_AngleCWT}
\end{figure}

Fig.~\ref{fig_AngleCWT} shows  values of $\theta^\ast$ of a wetting transition as a function of $\phi_S$ observed experimentally. Also included are the theoretical calculations with  Eq.(\ref{eq_rec}) with the value $a = 3.1 \pm 1.3$ obtained from fitting. Note that this value is close to previously obtained by a different method.~\cite{dubov:2014_2}  A general conclusion from this plot is that there are two distinct regimes of the wetting transition depending on $\phi_S$. In the first regime the elastic deformation of the contact line on the strong defects (see inset in Fig.~\ref{fig_AngleCWT}) reaches its maximum providing the receding angle above $90^\circ$. In the second regime the maximum deformation and a corresponding receding angle cannot be attained since when $\theta^\ast$ reaches the value of $90^\circ$ the Cassie state becomes thermodynamically unstable. By using Eq.(\ref{eq_rec}) we can now define a value of a solid fraction, $\phi_S^\ast$, which determines a criterion of two regimes of a wetting transition for a striped surface
\begin{equation}
1 = \phi_{S}^\ast (1 + \cos\theta_{r})-\frac{2a+1}{2}(\phi_S^\ast)^2\ln\phi_S^\ast,
\label{eq_regime}
\end{equation}
which for our surfaces gives $\phi^\ast_S\simeq 0.4$.

\begin{figure}[h]
  \centering
  \includegraphics[width=8cm]{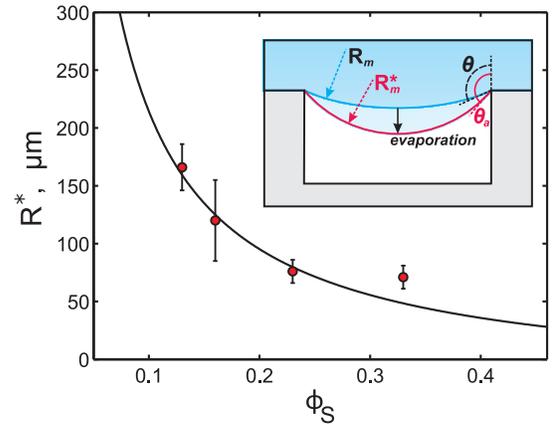}\\
  \caption{The critical radius of the drop provoking a transition as a function of $\phi_S$ for the surfaces with $d=5\mu$m. Circles show experimental data, the solid curve is the prediction of Eq.(\ref{eq_cwtradius}). Inset shows an evolution of a meniscus during evaporation.}
  \label{fig_R_CWT}
\end{figure}

%\textbf{Give a brief phrase  describing the purpose of next paragraph. What is our aim? The novelty and the strategy of a derivation.}

In the first regime, $\phi_S<\phi_S^\ast$, where a wetting transition commences at $\theta^\ast_r$ defined by Eq.(\ref{eq_rec}), the Cassie state is generally metastable according to Eq.(\ref{thitac}), so that the curvature of the evaporating drop is expected to provoke a transition. We have therefore measured the critical radius of the evaporating droplet, $R^\ast$, which leads to a transition, and the results are shown in Fig.~\ref{fig_R_CWT}. Let us now examine why a transition should take place. We first note that for our textures we have to rule our a mechanism related to touching the bottom part of texture by meniscus,~\cite{reyssat.m:2008} which would be typical for very dilute patterns. According to a criterion, $\left|\cos\theta_a\right|=4hw/(4h^2+w^2)$,~\cite{dubov:2013} such a scenario could be expected only for a texture with a twice larger $w$ than we use here, i.e. our textures are not dilute enough to expect such a scenario. Therefore, our transition is likely provoked by a violation of the pressure balance due to a bending of a meniscus between neighboring stripes.~\cite{dubov:2013,Tuteja2008} This balance can be written as $\Delta p (1-\phi_S)=\gamma \Lambda \left| \cos\theta \right|$,~\cite{dubov:2013,Antonini2014}
where $\Delta p$ is a Laplace pressure, $\Lambda$ is a density of contact line (i.e. a length of the triple contact line per unit area), and the instantaneous local contact angle $\theta$ is  on the edge of the groove (relative to a vertical (side) wall as seen in Fig.~\ref{fig_R_CWT}). The Laplace pressure under meniscus is $\Delta p =2\gamma/R=\gamma/R_m$, where $R_m$ is the radius of cylindrical meniscus, and the contact line density is $\Lambda=2/(w+d)=2\phi_S/d$. Therefore, for stripes $(1-\phi_S)/R=\left| \cos\theta \right|/(w+d)$. During the evaporation both the meniscus curvature and the instantaneous local contact angle (relative to a side wall), $\theta$, increase. The transition then occurs when the local contact angle reaches its maximum attainable value $\theta=\theta_a$. Thus the critical drop radius is
\begin{equation}
R^\ast=\frac{d(1-\phi_S)}{\phi_S \left| \cos\theta_a \right|}.
\label{eq_cwtradius}
\end{equation}
Note that it depends not only on $\phi_S$, but also on the spacing between stripes $d$ and properties of a material. The predictions of Eq.(\ref{eq_cwtradius}) are included in Fig.~\ref{fig_R_CWT} and are in excellent agreement with experimental data. Thus, Eq.(\ref{eq_cwtradius}) taken together with Eq.(\ref{eq_rec}) defines a criterion for a wetting transition on a relatively dilute stripes. We note that the strategy of our analysis for a striped SH surface is similar to previously suggested for isotropic array of pillars,~\cite{dubov:2013} although the relevance to $\cos\theta_r^{\ast}$, which depends on the contact line elasticity, remained obscure before.

\begin{figure}[h]
  \centering
  \includegraphics[width=4cm]{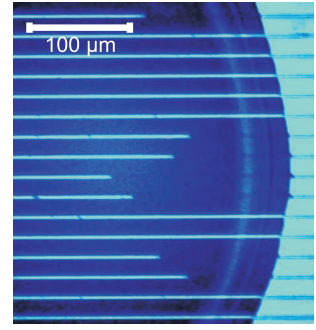}\\
  \caption{Experimental recording of the imbibition of the grooves during a wetting transition on the texture with $\phi_S=0.86$. }  \label{fig_jumps}
\end{figure}

In the second regime, $\phi_S>\phi_S^\ast$, the stage of a constant $\theta^\ast_r$ is not reached owing to the thermodynamic limitations. The surface becomes no longer effectively hydrophobic on the first stage of the evaporation. In this situation the values of $R^\ast$ were found to vary significantly between different measurements and also depend on the initial drop volume. These suggest that the transition is not really controlled by the curvature of the drop. The optical observations show that when $\theta^\ast$ reaches 90$^\circ$, the liquid starts to impregnate the grooves (see Fig.~\ref{fig_jumps}). We remark and stress that the onset of the transition is local. It starts at the triple contact line, but closer to center of the droplet the Cassie state is kept. Similar observations have been mentioned before.~\cite{bahadur:2009,bormashenko:2009comment,bormashenko:2007_2} Our work thus allows one interpreting them in terms of the elastic energy of the deformed contact line. Note that our impregnation velocity is 10--20~$\mu$m/s, so that the impregnation transition time is of the order of 10~s, i.e. several orders of magnitude exceeds the impalement transition time in case of dilute patterns.

%\section{Conclusion}

\textbf{Conclusion.~--}
 We have studied the wetting transition on striped Cassie surfaces with a different
fraction of solid areas. Our experiment demonstrated that that depending on the fraction of the solid area, and, therefore, elastic energy of the deformed
contact line, two qualitatively different scenarios of the transition occur. For relatively dilute stripes we have observed the sudden impalement transition controlled by a curvature of an evaporating drop. For dense stripes the slow impregnation transition  commences when the Cassie surface becomes effectively hydrophilic due to a decrease in an apparent receding angle. This transition represents the impregnation of the grooves from the triple contact line towards the drop center. We provide theoretical arguments, which allowed us to quantify and control both types of the wetting transition. Our interpretation thus allows one to use different recipes for avoiding this detrimental effect in experimental studies and various applications.

%\section{Acknowledgements}

\textbf{Acknowledgements.~--}
This research was partly supported by the Dynasty Foundation (grant of A.L.Dubov).

\bibliographystyle{aip}

%\bibliography{CWT_bib}% Produces the bibliography via BibTeX.

\end{document}